# Valley-dependent wavepacket self-rotation and Zitterbewegung in symmetry-broken honeycomb lattices

Xiuying Liu[1,+], Frane Lunić[2,+], Daohong Song[1,3], Zhixuan Dai[1], Shiqi Xia[1], Liqin Tang[1], Jingjun Xu[1,3], Zhigang Chen[1,3,4], and Hrvoje Buljan[1,2]

[1]*The MOE Key Laboratory of Weak-Light Nonlinear Photonics, TEDA Applied Physics Institute and School of Physics, Nankai University, Tianjin 300457, China*
[2]*Department of Physics, Faculty of Science, University of Zagreb, Bijenička cesta 32, 10000 Zagreb, Croatia*
[3]*Collaborative Innovation Center of Extreme Optics, Shanxi University, Taiyuan, Shanxi 030006, People's Republic of China*
[4]*Department of Physics and Astronomy, San Francisco State University, San Francisco, California 94132, USA*
songdaohong@nankai.edu.cn, hbuljan@phy.hr, zgchen@nankai.edu.cn
[+]These authors made equal contributions.

## Abstract

The toolbox quantities used for manipulating the flow of light include typically amplitude, phase, and polarization. Pseudospins, such as those arising from valley degrees of freedom in photonic structures, have recently emerged as an excellent candidate for this toolbox, in parallel with rapid development of spintronics and valleytronics in condensed-matter physics. Here, by employing symmetry-broken honeycomb photonic lattices, we demonstrate valley-dependent wavepacket self-rotation manifested in spiraling intensity patterns, which occurs without any initial orbital angular momentum. Theoretically, we show that such wavepacket self-rotation is induced by the Berry phase and results in Zitterbewegung oscillations. The "center-of-mass" of the wavepacket oscillates at a gap-dependent frequency, while the helicity of self-rotation is valley-dependent, that is, correlated with the Berry curvature. Our results lead to new understanding of the venerable Zitterbewegung phenomenon from the perspective of topology and are readily applicable on other platforms such as two-dimensional Dirac materials and ultracold atoms.

# Introduction

Electric charge is the key quantity for controlling signals in conventional electronics and semiconductor devices. However, advances in manipulating spin and valley degrees of freedom have reshaped the traditional perspective, leading to the development of spintronics (*1*) and valleytronics (*2*). Amplitude, phase, and polarization are the key quantities of usual recipes for controlling the flow of light. However, the understanding and development of optical spin–orbit interactions (*3*), photonic pseudospins (*4*), and valley degrees of freedom (*5-12*) have offered us new knobs that can be used for manipulation of light in photonic structures, in analogy with parallel advances in electric systems. In condensed matter structures, a local minimum in the conduction band or local maximum in the valence band is referred to as a valley (*2*). Among the most studied examples in electronics (*2*) and photonics (*13, 14*) are the two inequivalent valleys with degenerate energies in the honeycomb lattices (e.g., in graphene), located at the *K* and *K′* points in the Brillouin zone, which are furnished with the nontrivial Berry phase winding.

The pioneering achievements exploiting valley degrees of freedom in photonics include, for example, the prediction (*5, 6*) and experimental demonstration (*8*) of photonic valley-Hall topological insulators, topologically protected refraction of robust kink states in valley photonic crystals (*7*), topological valley Hall edge states (*9*), and spin and valley polarized one-way Klein tunneling (*11*). Photonic valley systems can be implemented at telecommunication and terahertz wavelengths on a silicon platform (*12, 15*), on subwavelength scales on plasmonic platforms (*8, 16, 17*), and they can be used for the development of topological lasers (*18-21*), which opens the possibilities for many applications. Besides electromagnetic waves, valley topological materials have been used for manipulation of other waves such as sound waves (*22*) and elastic waves (*23*). All these exemplary successes unequivocally point at the need and importance of discovering valley-dependent wave phenomena, for both fundamental understanding and advanced applications.

To this end, it is important to understand the behavior of physical quantities that distinguish different valleys. In photonics, the concept of Berry curvature is typically employed, and it points in opposite directions at *K*- and *K′*-valleys in a symmetry-broken photonic honeycomb-lattice (HCL) (e.g., see (*14*)). Besides the Berry curvature, in electronic systems, the electron magnetic moment can also distinguish between valleys (*24, 25*). The magnetic moment occurs from the self-rotating electric wavepacket (*24-26*), which is virtually impossible to directly observe with electrons.

Here we study valley-dependent propagation of light in an inversion-symmetry-broken photonic HCL. We establish the lattice by employing a direct laser-writing technique (*27*), and we demonstrate experimentally and numerically the valley-dependent helicity in spiraling intensity patterns related to wavepacket self-rotation. Specifically, we show that, by selective excitation of the valleys in a gapped HCL, a probe beam undergoes distinct spiraling during propagation through the lattice, characterized by its helical intensity pattern and "center-of-mass" oscillation, even though no initial orbital angular momentum is involved. We theoretically demonstrate that the observed phenomenon dwells upon the existence of the Berry phase (*28*), leading to the fundamental phenomenon of Zitterbewegung, first introduced by Schrödinger (*29*) in the context of relativistic electrons. We find that the helicity of Zitterbewegung in our system is a valley-dependent quantity.

The Zitterbewegung was studied in attempts to provide a deeper understanding of the electron spin (*30, 31*) and even to interpret some aspects of quantum mechanics (*32*), but the Zitterbewegung of electrons in vacuum has never been observed owing to its inherent ultra-small amplitude and ultra-high frequency. However, electrons in Bloch bands of some materials are driven by equations analogous to the relativistic Dirac equation; for example, Zitterbewegung of electrons was predicted to occur in semiconductor quantum wells (*33*). In a full analogy, Zitterbewegung was also predicted with ultracold atoms in optical lattices (*34*) and with photons in two-dimensional (2D) photonic crystals (*35*). Experimental observation of Zitterbewegung-like phenomena was, however, mostly limited to 1D domain in systems including trapped ions (*36*), photonic lattices (*37*), and Bose-Einstein condensates (*38, 39*), or to surface acoustic waves in an integrated phononic graphene (*40*). Traditionally, the Zitterbewegung is interpreted in terms of interference of positive and negative energy states, which in periodic systems amounts to interference of Bloch modes from two different bands. In this work, we show theoretically that the Zitterbewegung can be interpreted via interference between the incident non-vortex beam component and the vortex component arising from the universal momentum-to-real space mapping mechanism, which inherently has a topological origin (*41*). Thus, we provide a different perspective about the Zitterbewegung phenomenon, which gives rise to a simpler visualization than the original interpretation involving positive and negative energy states.

## Results
### Experimental results and numerical simulations

We study light propagation in (2+1)D photonic lattices, which in the paraxial approximation is governed by the Schrödinger-like equation (e.g. see (*14*) and Refs. therein),

$$i\frac{\partial \Psi}{\partial z} = -\frac{1}{2k_0}\nabla^2 \Psi - \frac{k_0 \delta n(x,y)}{n_0}\Psi(x,y,z). \tag{1}$$

Here, $\Psi(x,y,z)$ is the complex amplitude of the electric field, $k_0$ is the wave number in the medium, $n_0$ is the background refractive index, and $\delta n(x,y)$ is the induced refractive-index changes forming the HCL with broken inversion-symmetry, as illustrated in Fig. 1A. Equation (1) is mathematically equivalent to the Schrödinger equation describing electrons in 2D quantum systems, with $z$ playing the role of time. The HCL is comprised of two sublattices (*A* and *B*), and the inversion-symmetry breaking is achieved with a refraction index offset between the sublattices, see Fig. 1A. In $k$-space, the HCL has two distinct valleys located at the $K$ and $K'$ points in the Brillouin zone (they are also referred to as Dirac points), as illustrated in Fig. 1B. In the vicinity of Dirac points, the band structure is described by $\pm\sqrt{k^2 + m^2}$ ($k$ is the magnitude of the wavevector with origin at the Dirac point, and $m$ is the effective mass determined by band dispersion), and the wave dynamics is approximately described by the 2D massive Dirac equation (see theoretical analysis below). The size of the band gap is $2m$, which directly corresponds to (and thus can be controlled by) the refraction index offset between the two sublattices (see Figs. 1(A, B)).

Our main finding is sketched in Fig. 1B. The probe beam which is formed by interfering three broad Gaussian beams excites the modes in the vicinity of three equivalent $K$-points (or $K'$-points) in momentum space, i.e., the modes in one valley, with both sublattices equally excited in real space. The output beam exhibits self-rotation during propagation through the HCL, which has a spiraling intensity pattern with the helicity depending on the valley ($K$ or $K'$) that is initially excited. It will be shown below that this spiraling self-rotating motion is attributed to a root of the Zitterbewegung of the wavepacket, identified through the rotation of its "center-of-mass" (COM).

In Figs. 1(C-G), we show numerical results of the output patterns of the probe beam at different propagation distance in the inversion-symmetry-broken HCL by solving Eq. (1), with the refractive index offset between the sublattices set by the ratio $n_A : n_B = 1.2:1$, exciting only the $K$ valley. The parameters used in the simulations correspond to that of the experiment: $n_0 = 2.35$ for the SBN:61 crystal, $k_0 = 2\pi n_0 / \lambda$ and $\lambda = 488$ nm, the lattice constant is 16 μm (i.e., the distance between nearest neighboring sites is 9 μm), and the maximal index change (depth of the lattice) is about $1.3 \times 10^{-4}$. The overall envelope of the probe beam is Gaussian-like (Fig. 1D), but with a triangular lattice structure (due to three-beam interference) at $z = 0$ that can be positioned to excite one or both sublattices. In simulations displayed in Figs. 1(C-G), the probe beam excites the middle points between the ***A*** and ***B*** sublattices, i.e., both sublattices are equally excited. We find that the output wavepacket exhibits self-rotation during propagation (see the supplementary video), and the initially symmetric probe beam evolves into an asymmetric spiraling intensity pattern as displayed in Figs. 1(E-G). It expands during propagation because of diffraction, whereas the spiral helicity and the direction of rotation are valley-dependent. In Fig. 1C, the dynamical evolution of the beam's COM is plotted in 3D, showing spiral-like Zitterbewegung oscillation (in the plot we subtracted the drift which standardly occurs alongside Zitterbewegung phenomenon for better visualization).

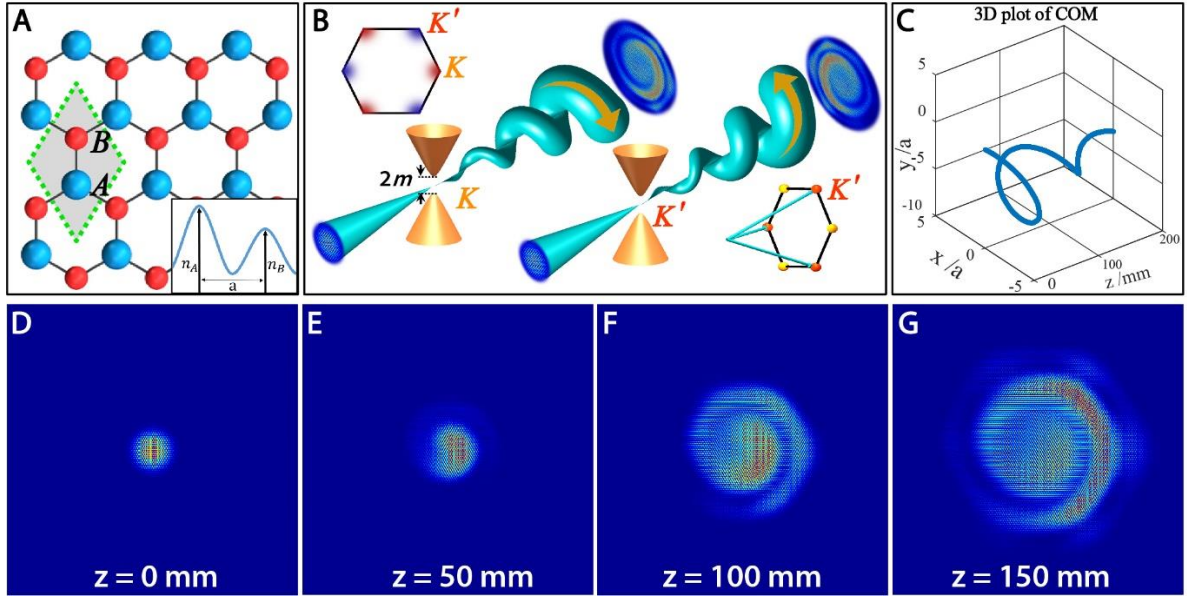

**Fig. 1. Valley-dependent wavepacket self-rotation in a symmetry-broken HCL.** (**A**) Illustration of an inversion-symmetry-broken HCL consisting of **A** and **B** sublattices. The inset sketches the refractive-index offset ($n_A > n_B$ is shown for example). (**B**) Illustration of wavepacket self-rotation when the modes in the vicinity of the $K$-valley (or $K'$-valley) are excited, showing spiraling intensity patterns with valley-dependent helicity. Top inset shows the valley locations at the edges of the Brillouin zone in $k$-space; the Berry curvature is opposite at two inequivalent valleys (sketched with red and blue colors). The gap size ($2m$) depends on the index offset ($n_A - n_B$). Bottom inset shows the scheme when three $K'$ valleys are simultaneously excited. (**C-G**) Spiraling COM (C) and intensity patterns obtained numerically at different propagation distances (E-G) indicate self-rotation of the wavepacket. The probe at $z = 0$ shown in (D) has a Gaussian envelope with no initial orbital angular momentum—see Supplementary video file.

Next, we present corresponding experimental results obtained in an HCL established in a 20-mm-long nonlinear crystal by a cw-laser-writing method (*27*). Instead of using a single Gaussian beam for writing, here the two sublattices are separately written and controlled by a triangular lattice pattern. The refractive-index difference of the two sublattices $n_A : n_B$ is readily tuned by the writing time for each sublattice (See Methods). A typical example of experimentally generated symmetry-broken HCL with $n_A > n_B$ is shown in Fig. 2A. As in simulation, the probe beam is a truncated triangular lattice pattern formed by interfering three broad Gaussian beams (see Fig. 2B) with their wavevectors matched to the three $K$- or $K'$-points. In real space, we excite both sublattices with equal amplitude and phase by positioning the probe beam at middle points between the two sublattices. The observed intensity patterns of the probe beam at the lattice output under different excitation conditions are shown in the top panels of Figs. 2(C-F), with corresponding numerical simulation results plotted in the bottom panels.

When the input beam excites the $K$-valley with the refractive-index offset between sublattices such that $n_A > n_B$, the beam evolves into a spiraling pattern (Fig. 2C). The helicity of the spiraling pattern and therefore the rotation direction of the output beam is reversed if the offset is changed to be $n_A < n_B$ (Fig. 2D). As we shall show theoretically below, such spiraling intensity pattern is related to the circular motion of the COM of the wavepacket and the Berry-phase-mediated Zitterbewegung. We emphasize that the rotation can only be realized when the inversion symmetry of the HCL is broken and the gap opens; for comparison, when $n_A = n_B$, the output pattern exhibits conical diffraction (*42*) rather than a spiraling pattern (Fig. 2E) under the equal excitation condition. Importantly, we experimentally demonstrate that the rotation direction depends on the valley degree of freedom. If we excite the $K'$ valley instead of the $K$ valley, while keeping all other conditions unchanged, we observe that the spiraling direction (i.e., helicity) of the intensity pattern is reversed; this can be seen by comparing the experimental results shown in Figs. 2C and 2F. These observations are corroborated by numerical beam propagation simulations using Eq. (1), which is shown in the bottom panels of Fig. 2. We point out that altering the helicity of the spiraling pattern by reversing the index offset between the two sublattices (Fig. 2C vs. Fig. 2D) is fully equivalent to altering the helicity via exciting different valleys (Fig. 2C vs. Fig. 2F). In both cases, the helicity of the spiraling pattern is correlated with the direction of the Berry curvature around the gapped Dirac cone. In other words, the spiraling intensity is a valley contrasting quantity, analogous to the orbital momentum of electrons in condensed matter systems (*24-26*), manifested when the inversion symmetry is broken.

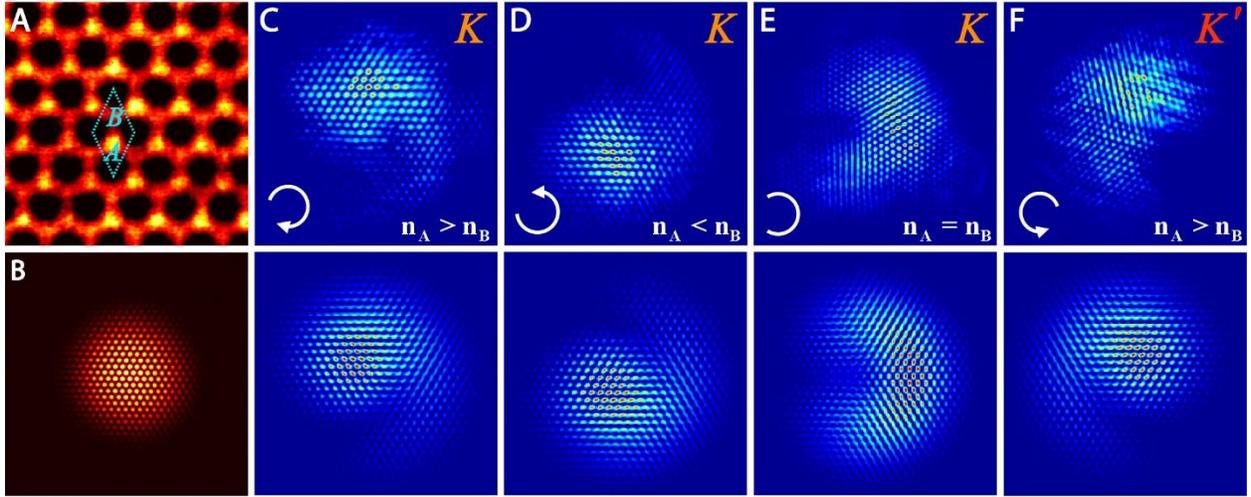

**Fig. 2. Experimental and numerical results demonstrating valley-dependent wavepacket self-rotation.** (**A**) Zoom-in image of a laser-written HCL with broken-inversion-symmetry; in this plot, $n_A > n_B$, corresponding to Fig. 1(A). (**B**) Input triangular lattice pattern used in experiment as the probe beam. (**C-F**) Experimental (top row) and numerical (bottom row) results of output intensity patterns for different excitation conditions: (C-E) Results obtained under $K$-valley excitation where the index ratio is (C) $n_A:n_B = 1.2:1$, (D) $n_A:n_B = 1:1.2$, and (E) $n_A:n_B = 1:1$; (F) Result obtained under $K'$-valley excitation with $n_A:n_B = 1.2:1$. Note that the helicities of the spiraling patterns in (C and F) (as well as in C and D) are in opposite directions, as illustrated by curved arrows.

## Theoretical analysis

For excitations in the vicinity of the $K$-valley, Eq. (1) is approximated by $i\frac{\partial \psi}{\partial z} = H\psi$, where the Hamiltonian (in $k$-space) is an effective 2D massive Dirac equation:

$$H = \kappa(\sigma_x k_x + \sigma_y k_y) + \sigma_z m = \begin{pmatrix} m & \kappa(k_x - ik_y) \\ \kappa(k_x + ik_y) & -m \end{pmatrix} = \begin{pmatrix} m & \kappa k e^{-i\varphi_k} \\ \kappa k e^{i\varphi_k} & -m \end{pmatrix}, \quad (2)$$

where $\sigma_i$ are the Pauli matrices. The coefficient $\kappa$ depends on the coupling strength between adjacent waveguides in the lattice (e.g., see (*25*)). Without any loss of generality, we set $\kappa = 1$ in all analytical expressions, because they can be rescaled to any value of $\kappa$ with the substitution

$k \to \kappa k$. The complex amplitude of the electric field $\psi = \begin{pmatrix} \psi_{\frac{1}{2}} \\ \psi_{-\frac{1}{2}} \end{pmatrix}$ is a two-component spinor, because the HCL has two sublattices. Pseudospin components $\psi_{\frac{1}{2}}$ and $\psi_{-\frac{1}{2}}$ describe the field amplitudes in the *A* and *B* sublattices (e.g., see (*41*)). Dynamics around the *K'*-valley is described analogously with the substitution $k_x \to -k_x$ in Eq. (2) (*25*). The geometry of the eigenmodes gives rise to the Berry curvature which is in opposite directions at the *K* and *K'* points (*14, 25, 26*); see Fig. 1B.

We are interested in the dynamics from an axially symmetric initial excitation,

$$\psi(r, \varphi_r, z = 0) = \psi_0 \sqrt{I_0(r)} = \int d^2k\, \psi_0 f(k) e^{i\mathbf{k}\cdot\mathbf{r}}, \tag{3}$$

where we have introduced radial coordinates ($x = r\cos\varphi_r$ and $y = r\sin\varphi_r$), and $\psi_0 = \begin{pmatrix} \cos\theta\, e^{i\alpha} \\ \sin\theta \end{pmatrix}$ is the most general initial spinor; $\alpha$ is the relative phase between the fields in the sublattices at $z = 0$, and $\theta$ determines the amplitude in each sublattice. After a straightforward calculation one finds

$$\psi(r, \varphi_r, z) = \begin{pmatrix} \psi_{\frac{1}{2}}(r, \varphi_r, z) \\ \psi_{-\frac{1}{2}}(r, \varphi_r, z) \end{pmatrix} = \begin{pmatrix} g_{\frac{1}{2},0}(r,z) + g_{\frac{1}{2},-1}(r,z) e^{-i\varphi_r} \\ g_{-\frac{1}{2},+1}(r,z) e^{i\varphi_r} + g_{-\frac{1}{2},0}(r,z) \end{pmatrix}, \tag{4}$$

where $z = 13/\kappa k_0$, and the *g*-functions can be expressed as integrals in *k*-space (see Supplementary Material). In Fig. 3A we plot the spiraling intensity pattern $\left|\psi_{\frac{1}{2}}(r, \varphi_r, z)\right|^2$ obtained with the Hamiltonian in Eq. (2); the envelope of the initial excitation is Gaussian, $f(k) = \exp(-k^2/k_0^2)$, and both sublattices are equally excited with same phase, $\psi_0 = \begin{pmatrix} 1 \\ 1 \end{pmatrix}$. The mass term is $m = 0.6\kappa k_0$, which determines the gap size. It is evident that the spiraling intensity pattern obtained with the "low-energy" Hamiltonian Eq. (2) agrees with those obtained from numerical simulations of the Schrödinger equation (1) as well as from experiments shown in Fig. (2).

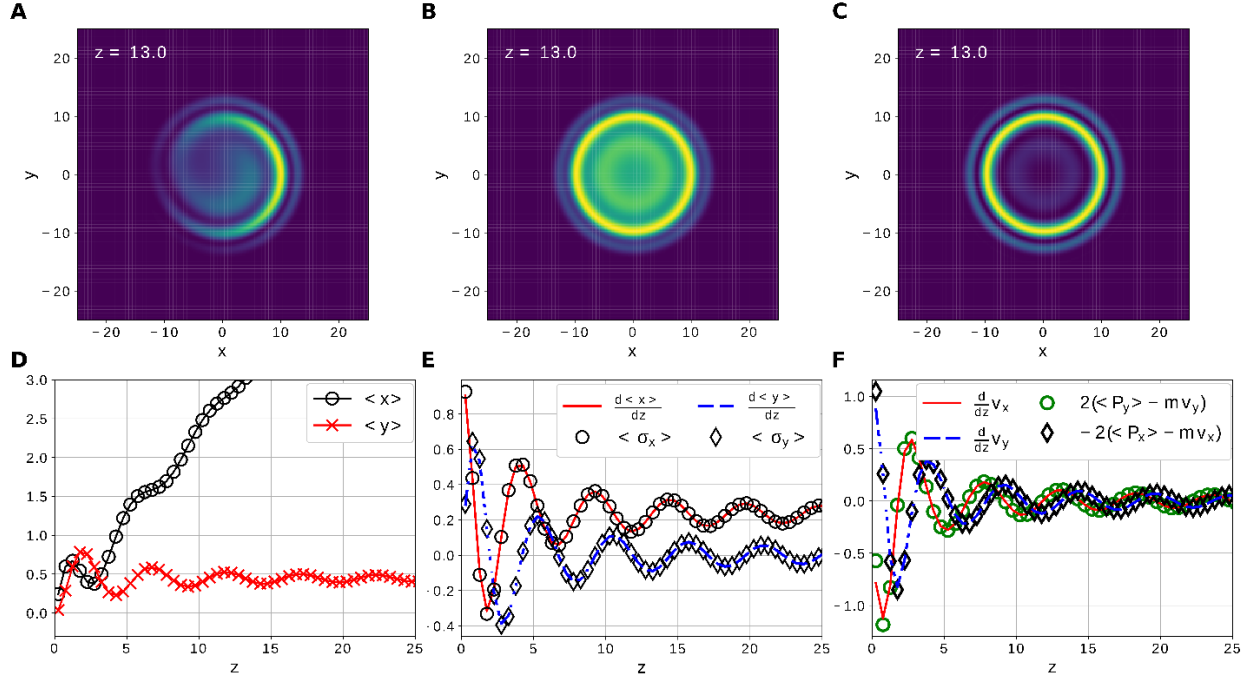

**Fig. 3. Theoretical analysis of wavepacket self-rotation from Dirac equation.** Top panels are the spinor components of the intensity structure of the spiraling beam, and bottom panels show the motion of its "center-of-mass (COM)". In the figure, $z$ is in units $(\kappa k_0)^{-1}$, $x$ and $y$ are in units $k_0^{-1}$. (**A**) Intensity structure of the pseudospin component $\psi_{\frac{1}{2}}(r,\varphi_r,z)$, (**B**) the non-vortex component $\left|g_{\frac{1}{2},0}(r,z)\right|^2$, and (**C**) the vortex component $\left|g_{\frac{1}{2},-1}(r,z)\right|^2$. (**D**) The position of the COM of the wavepacket (average values of $x$ and $y$) as functions of $z$. (**E**) Propagation of the velocity components of the COM, and the (identical) expectation values $\langle\sigma_x\rangle$ and $\langle\sigma_y\rangle$. (**F**) Propagation of the acceleration components and numerical verification of Eq. (9). See text for details.

It is important to note from Eq. (4) that each spinor component is a superposition of a non-vortex (Gaussian-like) amplitude and a vortex field amplitude. To explain the spiraling pattern observed in our experiments, we calculate the intensity in the pseudospin components:

$$\left|\psi_{\frac{1}{2}}\right|^2 = \left|g_{\frac{1}{2},0}(r,z)\right|^2 + \left|g_{\frac{1}{2},-1}(r,z)\right|^2 + 2\left|g_{\frac{1}{2},0}\right|\left|g_{\frac{1}{2},-1}\right|\cos(-\mathrm{Arg}\,g_{\frac{1}{2},0}(r,z) + \mathrm{Arg}\,g_{\frac{1}{2},-1}(r,z) - \varphi_r), \quad (5)$$

and equivalently for the other pseudospin component. The last term describes the interference between the vortex and non-vortex field amplitudes, which depends on their relative phase. The intensities of the non-vortex term $\left|g_{\frac{1}{2},0}(r,z)\right|^2$ and the vortex term $\left|g_{\frac{1}{2},-1}(r,z)\right|^2$ are radially symmetric, as shown in Figs. 3B and 3C. Therefore, the spiraling pattern must arise from the interference. The interference term has a maximum when

$$\varphi_r = -\mathrm{Arg}\,g_{\frac{1}{2},0}(r,z) + \mathrm{Arg}\,g_{\frac{1}{2},-1}(r,z) \quad (\text{modulo } 2\pi). \quad (6)$$

When $-\mathrm{Arg}\,g_{\frac{1}{2},0}(r,z) + \mathrm{Arg}\,g_{\frac{1}{2},-1}(r,z)$ is monotonically increasing (or decreasing) with $r$, the function implicitly given in Eq. (6) is a spiral in the $(r,\varphi_r)$-plane; the spiral helicity depends on whether the r.h.s. in Eq. (6) decreases or increases. Evidently, the spiraling self-rotating pattern arises from the interference of the vortex and the non-vortex components.

We now present the theory for the wavepacket self-rotation and Zitterbewegung phenomenon in our system. Dynamics of the COM of the wavepacket $\mathbf{r}_C = x_C \hat{x} + y_C \hat{y}$ is given by

$$\mathbf{r}_C(z) = \langle \mathbf{r} \rangle = \int \psi^+(r, \varphi_r, z) \mathbf{r} \psi(r, \varphi_r, z) da, \tag{7}$$

where $\mathbf{r} = x\hat{x} + y\hat{y}$, and $da = r dr d\varphi_r$ is the infinitesimal area element. It can be understood by observing the velocity of the COM,

$$\mathbf{v}_C = \frac{d\mathbf{r}_C}{dz} = \int \psi^+(r, \varphi_r, z) i[H, \mathbf{r}] \psi(r, \varphi_r, z) da = \langle \sigma_x \rangle \hat{x} + \langle \sigma_y \rangle \hat{y}, \tag{8}$$

and its acceleration,

$$\frac{d\mathbf{v}_C}{dz} = -2 \hat{z} \times \mathbf{P} + 2m \hat{z} \times \mathbf{v}_C; \tag{9}$$

here we have introduced vector $\mathbf{P} = \langle k_x \sigma_z \rangle \hat{x} + \langle k_y \sigma_z \rangle \hat{y}$ (see Supplementary Material for the derivation). Calculated results from Eqs. (7-9) are illustrated in Figs. 3(D-F).

The second term in Eq. (9) is the Zitterbewegung term; it corresponds to the oscillations of the COM with frequency $2m$ (the size of the spectral gap). Oscillations are clearly visible in all Figs. 3(D-F). Moreover, it is evident from Eq. (9) that the helicity of Zitterbewegung oscillations depends on the sign of $m$, which corroborates our experimental findings. The first term in Eq. (9) yields the drift of the COM of the wavepacket, visible in Fig. 3D, which is an expected feature of the Zitterbewegung effect (e.g., see (*33, 43*)). The direction of the drift depends on the initial conditions. More specifically, the expectation value of the pseudospin operator $\boldsymbol{\sigma} = \hat{x}\sigma_x + \hat{y}\sigma_y$ at $z = 0$ sets the direction of the initial velocity of the COM (see Fig. 3E). Such drifting of the COM is also observed in our numerical simulations using Eq. (1). We note that for better visualization of the spiraling dynamics, we did not include the drift when plotting Fig. 1C.

The components of the vector $\mathbf{P}$ are interpreted as the difference of the expectation value of the momentum between the pseudospin-up and -down components, that is, the difference of the momentum between the two sublattices. The acceleration of the COM in the $x$-direction is proportional to $P_y$, which can be therefore interpreted as a pseudo-force exerted in the COM. From the example shown in Fig. 3F, we see that this pseudo-force $\mathbf{P}$ oscillates around zero. Thus, it induces some oscillations, which should be distinguished from the Zitterbewegung term. Our calculations indicate that the circular Zitterbewegung motion in symmetry-broken HCLs exists only when $m$ is nonzero and thus the gap opens, which is in agreement with the Zitterbewegung of electrons (*29*). Yet, our finding is in contradistinction with similar oscillations that were called Zitterbewegung in gapless honeycomb lattice systems (*35, 40, 44*).

## Discussion
The theory of the Zitterbewegung has been addressed in numerous papers (*30, 32-35, 45-47*). The Zitterbewegung effect was originally associated with circular motion of electrons in 3D space (*29, 30*), but such motion has never been observed. Here, we focus on the novel aspects of this phenomenon using optical wavepacket in 2D photonic lattices. We discuss connection between the experimentally observed valley-dependent spiraling intensity pattern (related to self-rotation of the wavepacket) and the Zitterbewegung phenomenon. This leads to a novel interpretation of the phenomenon, and sheds light on the role played by the Berry phase.

First, we mention a seemingly unrelated simple example. Considering two coupled single-mode waveguides, the coupled system has a symmetric and an anti-symmetric eigenmode, $\frac{1}{\sqrt{2}}(u_L \pm u_R)$, with two propagation constants (eigenvalues) whose difference depends on the strength of the coupling (here the letter $L$ stands for the left waveguide, and $R$ for the right waveguide). By launching a beam, for example, into the left waveguide, both modes will be excited and they will undergo beating; the field amplitude will thus jump from the left to the right waveguide and back and forth, with the frequency given by the coupling strength. The COM of the beam will oscillate at this frequency.

The very same mechanism, albeit a bit more complicated, leads to Zitterbewegung in our 2D system. First, we excite both sublattices of the HCL equally and with the same phase ($\psi_0 = \begin{pmatrix} \cos\theta\, e^{i\alpha} \\ \sin\theta \end{pmatrix} = \begin{pmatrix} 1 \\ 1 \end{pmatrix}$). The envelope of the initial excitation is Gaussian-like with azimuthal symmetry. In experiments and numerical simulations, the intensity fine structure under this envelope is a triangular lattice (it allows tuning the excitation of the two sublattices). In "low-energy" theory Eq. (2), this means that the continuous field amplitudes $\psi_{\frac{1}{2}}(r, \varphi_r, z = 0)$ and $\psi_{-\frac{1}{2}}(r, \varphi_r, z = 0)$ are independent of the azimuthal angle $\varphi_r$. Because of the nontrivial Berry phase winding around the Dirac points, that is, the topology of the system, a vortex beam component (with $\varphi_r$ dependent amplitude) will dynamically emerge. (The underlying universal mechanism which maps the topological singularity (vortex) from the $k$-space to the real space was discovered recently (*41*)). As such, a single pseudospin component is furnished with both the non-vortex and the vortex beam components, which naturally interfere. It is demonstrated in the previous section and shown in Fig. 3 that without the interference of these two components, the intensity pattern of the beam remains its azimuthal symmetry. The shape of the interference fringes depends on the evolution of the phase fronts of these two components, i.e., on $\mathrm{Arg}\, g_{\frac{1}{2},0}(r,z) - \mathrm{Arg}\, g_{\frac{1}{2},-1}(r,z)$, which yields a spiraling self-rotating pattern (see Fig. 3). This rotation breaks the azimuthal symmetry of the initial beam and leads to oscillation of the COM of the beam $\mathbf{r}_C(z)$, in analogy to the two-mode beating discussed above. This alternative interpretation of the Zitterbewegung oscillations is perhaps more easily visualized than the original one invoking interference between positive and negative energy states. Both interpretations are correct, however, ours gives a simple picture for the circular oscillations of the COM associated with wavepacket self-rotation.

Second, without the gap, there is no Zitterbewegung [see Eq. (9)]. This means that the gap is crucial for the existence of radial dependence of the phase fronts $\mathrm{Arg}\, g_{\frac{1}{2},0}(r,z) - \mathrm{Arg}\, g_{\frac{1}{2},-1}(r,z)$ that yields the spiraling intensity patterns. This can be understood because evolving phase fronts correspond to the dispersion curves. The dispersion curves drastically change when the gap opens, from the linear (conical) dispersion to the "parabolic" one. Third, the helicity of the spiraling self-rotating motion determines the helicity of the Zitterbewegung of the COM. Consequently, it is valley-dependent in the staggered HCLs.

Finally, the role of the Berry phase is crucial. The existence of the Berry phase at each valley is responsible for the existence of the momentum to real space mapping which produces a vortex component in the field, even though that the initial excitation beam is Gaussian-like. The connection between the Berry phase and Zitterbewegung has been analyzed previously in literature (*34*). These analyses relied on the fact that the COM of the beam can be expressed as $\langle \mathbf{r} \rangle = \int \tilde{\psi}^+(\mathbf{k}, z) i \nabla_\mathbf{k} \tilde{\psi}(\mathbf{k}, z) d^2 k$ in the momentum space representation of the field amplitude (*34, 46*). When $\tilde{\psi}(\mathbf{k}, z)$ is expressed in eigenmodes of the system, some terms in the expression for $\langle \mathbf{r} \rangle$ will contain the Berry connection $\mathbf{A}_n(\mathbf{k}) = i \psi_{n\mathbf{k}}^+ \nabla_\mathbf{k} \psi_{n\mathbf{k}}$ ; however, the terms corresponding to Zitterbewegung will be non-zero only if the interband matrix elements $i \psi_{-1\mathbf{k}}^+ \nabla_\mathbf{k} \psi_{1\mathbf{k}}$ and $i \psi_{1\mathbf{k}}^+ \nabla_\mathbf{k} \psi_{-1\mathbf{k}}$ are non-zero (see Supplementary Material for the derivation). These matrix elements take very similar form to that of the Berry connection, except that the operator $i \nabla_\mathbf{k}$ is evaluated between modes of different bands. This is consistent with our experimental setting where both bands are excited. Thus, we conclude that in our observations, the key role of the Berry phase is to generate the vortex term enabling its interference with the non-vortex component, and hence the Zitterbewegung. The direction of the Berry curvature sets the helicity of the spiraling pattern, and therefore the valley-dependence of the spiraling self-rotating wavepacket.

## Materials and Methods
### Experimental setup and scheme

In our experiment, the symmetry-breaking HCL is established in a nonlinear photorefractive crystal (SBN:61; dimensions: a × b × c = 5 × 20 × 5 mm$^3$ ) by using the cw-laser writing method established previously (*27*). Instead of using a single Gaussian beam for waveguide writing, here we employ a triangular lattice beam formed by three-beam interference. A collimated ordinarily-polarized laser beam at a wavelength of 488 nm illuminates a spatial light modulator (SLM) loaded with a programmable phase mask, which is transformed into a triangular lattice pattern after passing through a 4f system combined with a filter. Such a lattice beam remains invariant through the 20-mm-long crystal. The lattice-writing beam induces a triangular lattice due to the photorefractive self-focusing nonlinearity, as controlled by the beam power (4.8mW), the bias electric field (1.2×10$^5$ V/m), and the writing time. The HCL can be established by alternatively writing the two sublattices, as illustrated in the Fig. 4, taking advantage of the "memory effect" of the photorefractive crystal. In addition, since the value of nonlinear refractive index change is proportional to the writing time, we can readily tune the refractive index differences of the two sublattices by using different writing time for each sublattice. As such, the same triangular lattice beam induces different index changes for different sublattices, leading to the desired inversion-symmetry-broken HCL as examined by a broad (quasi-plane-wave) beam (see the insets in Fig 4). To selectively excite the HCL, an extraordinarily-polarized and truncated triangular lattice beam is sent into the lattice along the same optical path with the writing beam. However, the probe beam has a much smaller size that covers only several lattice sites, and its direction (with its three constituting components momentum-matched to three equivalent valleys) and launching position (for exciting both sublattices) can be precisely controlled by the SLM. To avoid the self-action of the probe beam due to nonlinearity, the intensity of the probe beam is set to be sufficiently low, so that it undergoes linear propagation. The output intensity patterns of the probe beam though the lattice are recorded by a CCD camera.

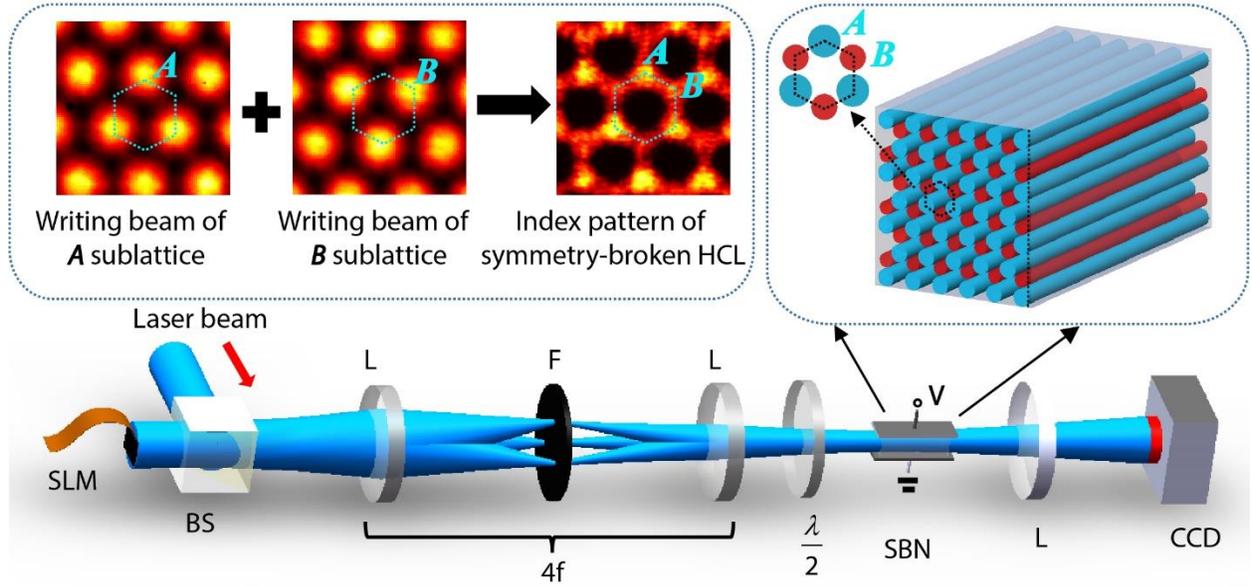

**Fig. 4. Experimental setup and scheme used for laser-writing the symmetry-broken HCL and for the observation of valley-dependent wavepacket self-rotation.** SLM, spatial light modulator; BS, beam splitter; L, lens; F, Filter; $\frac{\lambda}{2}$, half wave plate; SBN: strontium barium niobite crystal. The triangular lattice beam for alternatively writing the two sublattices and the superimposed lattice structure is shown in the top-left inset, and the 3D lattice structure through the crystal is illustrated in the top-right inset.

## Theory

Dynamics from the initial condition $\psi(r,\varphi_r,z=0) = \psi_0\sqrt{I_0(r)} = \int d^2k\,\psi_0 f(k)e^{i\mathbf{k}\cdot\mathbf{r}}$ is readily found by expanding into eigenmodes of the system. The eigenmodes of the Hamiltonian in Eq. (2), $\psi_{n\mathbf{k}}$, are given by $H\psi_{n\mathbf{k}} = \beta_{n\mathbf{k}}\psi_{n\mathbf{k}}$,

$$\psi_{n\mathbf{k}} = \frac{1}{\sqrt{N_{nk}}}\begin{pmatrix} \frac{m+\beta_{nk}}{ke^{i\varphi_k}} \\ 1 \end{pmatrix},\ N_{nk} = 2 + \frac{2m(m+\beta_{nk})}{k^2}.$$

where $\beta_{n\mathbf{k}} = n\sqrt{k^2+m^2}$; $n = \pm 1$ is the band number, and $\mathbf{k}$ is the wavevector with origin at the $K$-point. The propagating complex amplitude of the field is

$$\psi(r,\varphi_r,z) = \sum_n \int d^2k\,c_{n\mathbf{k}}\psi_{n\mathbf{k}}f(k)e^{i\mathbf{k}\cdot\mathbf{r}-i\beta_{nk}z}, \tag{3}$$

where the expansion coefficients are $c_{n\mathbf{k}} = \psi_{n\mathbf{k}}^+ \psi_0$. After a straightforward calculation one derives Eq. (4). The $g$-functions are expressed as integrals in $k$-space. The $z$-derivative of any operator $O$ is calculated via $\frac{dO}{dz} = i[H, O]$, which yields Eqs. (8) and (9). See Supplementary Material for details of the calculation.

**Acknowledgements**

This research is supported by the National Key R&D Program of China under Grant No. 2017YFA0303800, the National Natural Science Foundation (11922408, 91750204, 11674180), PCSIRT, and the 111 Project (No. B07013) in China. H.B. acknowledge support in part by the Croatian Science Foundation Grant No. IP-2016-06-5885 SynthMagIA, and the QuantiXLie Center of Excellence, a project co-financed by the Croatian Government and European Union through the European Regional Development Fund - the Competitiveness and Cohesion Operational Programme (Grant KK.01.1.1.01.0004).

# REFERENCES AND NOTES

1. J. F. Igor Žutić, S. Das Sarma, Spintronics: Fundamentals and applications. *Rev. Mod. Phys.* **76**, 323-410 (2004).
2. J. R. Schaibley, H. Yu, G. Clark, P. Rivera, J. S. Ross, K. L. Seyler, W. Yao, X. Xu, Valleytronics in 2D materials. *Nat. Rev. Mater.* **1**, 16055 (2016).
3. K. Y. Bliokh, F. J. Rodríguez-Fortuño, F. Nori, A. V. Zayats, Spin–orbit interactions of light. *Nat. Photonics* **9**, 796-808 (2015).
4. D. Song, V. Paltoglou, S. Liu, Y. Zhu, D. Gallardo, L. Tang, J. Xu, M. Ablowitz, N. K. Efremidis, Z. Chen, Unveiling pseudospin and angular momentum in photonic graphene. *Nat. Commun.* **6**, 6272 (2015).
5. T. Ma, G. Shvets, All-Si valley-Hall photonic topological insulator. *New J. Phys.* **18**, 025012 (2016).
6. J. W. Dong, X. D. Chen, H. Zhu, Y. Wang, X. Zhang, Valley photonic crystals for control of spin and topology. *Nat. Mater.* **16**, 298-302 (2017).
7. F. Gao, H. Xue, Z. Yang, K. Lai, Y. Yu, X. Lin, Y. Chong, G. Shvets, B. Zhang, Topologically protected refraction of robust kink states in valley photonic crystals. *Nat. Phys.* **14**, 140-144 (2017).
8. X. Wu, Y. Meng, J. Tian, Y. Huang, H. Xiang, D. Han, W. Wen, Direct observation of valley-polarized topological edge states in designer surface plasmon crystals. *Nat. Commun.* **8**, 1304 (2017).
9. J. Noh, S. Huang, K. P. Chen, M. C. Rechtsman, Observation of Photonic Topological Valley Hall Edge States. *Phys. Rev. Lett.* **120**, 063902 (2018).
10. X.-D. Chen, F.-L. Zhao, M. Chen, J.-W. Dong, Valley-contrasting physics in all-dielectric photonic crystals: Orbital angular momentum and topological propagation. *Phys. Rev. B* **96**, 020202 (2017).
11. X. Ni, D. Purtseladze, D. A. Smirnova, A. Slobozhanyuk, A. B. Khanikaev, Spin- and valley-polarized one-way Klein tunneling in photonic topological insulators. *Sci. Adv.* **4**, eaap8802 (2018).

12. M. I. Shalaev, W. Walasik, A. Tsukernik, Y. Xu, N. M. Litchinitser, Robust topologically protected transport in photonic crystals at telecommunication wavelengths. *Nat. Nanotechnol.* **14**, 31-34 (2019).
13. L. Lu, J. D. Joannopoulos, M. Soljačić, Topological photonics. *Nat. Photonics* **8**, 821-829 (2014).
14. T. Ozawa, H. M. Price, A. Amo, N. Goldman, M. Hafezi, L. Lu, M. C. Rechtsman, D. Schuster, J. Simon, O. Zilberberg, I. Carusotto, Topological photonics. *Rev. Mod. Phys.* **91**, 015006 (2019).
15. Y. Yang, Y. Yamagami, X. Yu, P. Pitchappa, J. Webber, B. Zhang, M. Fujita, T. Nagatsuma, R. Singh, Terahertz topological photonics for on-chip communication. *Nat. Photonics* **14**, 446-451 (2020).
16. M. Jung, Z. Fan, G. Shvets, Midinfrared Plasmonic Valleytronics in Metagate-Tuned Graphene. *Phys. Rev. Lett.* **121**, 086807 (2018).
17. M. Proctor, P. A. Huidobro, S. A. Maier, R. V. Craster, M. P. Makwana, Manipulating topological valley modes in plasmonic metasurfaces. *Nanophotonics* **9**, 657-665 (2020).
18. Y. Zeng, U. Chattopadhyay, B. Zhu, B. Qiang, J. Li, Y. Jin, L. Li, A. G. Davies, E. H. Linfield, B. Zhang, Y. Chong, Q. J. Wang, Electrically pumped topological laser with valley edge modes. *Nature* **578**, 246-250 (2020).
19. H. Zhong, Y. Li, D. Song, Y. V. Kartashov, Y. Zhang, Y. Zhang, Z. Chen, Topological Valley Hall Edge State Lasing. *LASER PHOTONICS REV* **14**, 2000001 (2020).
20. Y. Gong, S. Wong, A. J. Bennett, D. L. Huffaker, S. S. Oh, Topological Insulator Laser Using Valley-Hall Photonic Crystals. *ACS PHOTONICS* **7**, 2089-2097 (2020).
21. D. Smirnova, A. Tripathi, S. Kruk, M. S. Hwang, H. R. Kim, H. G. Park, Y. Kivshar, Room-temperature lasing from nanophotonic topological cavities. *Light.: Sci. Appl.* **9**, 127 (2020).
22. J. Lu, C. Qiu, L. Ye, X. Fan, Z. Liu, Observation of topological valley transport of sound in sonic crystals. *Nat. Phys.* **13**, 369–374 (2017).
23. M. Yan, J. Lu, F. Li, W. Deng, X. Huang, J. Ma, Z. Liu, On-chip valley topological materials for elastic wave manipulation. *Nat. Mater.* **17**, 993-998 (2018).
24. D. Xiao, W. Yao, Q. Niu, Valley-contrasting physics in graphene: magnetic moment and topological transport. *Phys. Rev. Lett.* **99**, 236809 (2007).
25. X. Xu, W. Yao, D. Xiao, T. F. Heinz, Spin and pseudospins in layered transition metal dichalcogenides. *Nat. Phys.* **10**, 343-350 (2014).
26. D. Xiao, M.-C. Chang, Q. Niu, Berry phase effects on electronic properties. *Rev. Mod. Phys.* **82**, 1959-2007 (2010).
27. S. Xia, A. Ramachandran, S. Xia, D. Li, X. Liu, L. Tang, Y. Hu, D. Song, J. Xu, D. Leykam, S. Flach, Z. Chen, Unconventional Flatband Line States in Photonic Lieb Lattices. *Phys. Rev. Lett.* **121**, 263902 (2018).
28. M. V. Berry, Quantal Phase Factors Accompanying Adiabatic Changes. *Proc. R. Soc. Lond A*, (1984).
29. E. Schrödinger, Über die kräftefreie Bewegung in der relativistischen Quantenmechanik. *Sitz. der Preuss. Akad. der Wiss. Phys.-Math. Kl.* **24**, 418-428 (1930).
30. K. Huang, On the Zitterbewegung of the Dirac Electron. *Am. J. Phys.* **20**, 479-484 (1952).
31. C.-P. Chuu, M.-C. Chang, Q. Niu, Semiclassical dynamics and transport of the Dirac spin. *Solid State Commun.* **150**, 533-537 (2010).
32. D. Hestenes, The Zitterhewegung Interpretation of Quantum Mechanics. *Found. Physics.* **20**, 1213-1232 (1990).
33. J. Schliemann, D. Loss, R. M. Westervelt, Zitterbewegung of electronic wave packets in III-V zinc-blende semiconductor quantum wells. *Phys. Rev. Lett.* **94**, 206801 (2005).
34. J. Y. Vaishnav, C. W. Clark, Observing Zitterbewegung with ultracold atoms. *Phys. Rev. Lett.* **100**, 153002 (2008).

35. X. Zhang, Observing Zitterbewegung for photons near the Dirac point of a two-dimensional photonic crystal. *Phys. Rev. Lett.* **100**, 113903 (2008).
36. R. Gerritsma, G. Kirchmair, F. Zahringer, E. Solano, R. Blatt, C. F. Roos, Quantum simulation of the Dirac equation. *Nature* **463**, 68-71 (2010).
37. F. Dreisow, M. Heinrich, R. Keil, A. Tunnermann, S. Nolte, S. Longhi, A. Szameit, Classical simulation of relativistic Zitterbewegung in photonic lattices. *Phys. Rev. Lett.* **105**, 143902 (2010).
38. L. J. LeBlanc, M. C. Beeler, K. Jiménez-García, A. R. Perry, S. Sugawa, R. A. Williams, I. B. Spielman, Direct observation of zitterbewegung in a Bose–Einstein condensate. *New J. Phys.* **15**, 073011 (2013).
39. C. Qu, C. Hamner, M. Gong, C. Zhang, P. Engels, Observation ofZitterbewegungin a spin-orbit-coupled Bose-Einstein condensate. *Phys. Rev. A* **88**, 021604(R) (2013).
40. S. Y. Yu, X. C. Sun, X. Ni, Q. Wang, X. J. Yan, C. He, X. P. Liu, L. Feng, M. H. Lu, Y. F. Chen, Surface phononic graphene. *Nat. Mater.* **15**, 1243-1247 (2016).
41. X. Liu, S. Xia, E. Jajtić, D. Song, D. Li, L. Tang, D. Leykam, J. Xu, H. Buljan, Z. Chen, Universal momentum-to-real-space mapping of topological singularities. *Nat. Commun.* **11**, 1586 (2020).
42. O. Peleg, G. Bartal, B. Freedman, O. Manela, M. Segev, D. N. Christodoulides, Conical diffraction and gap solitons in honeycomb photonic lattices. *Phys. Rev. Lett.* **98**, 103901 (2007).
43. G. Dávid, J. Cserti, General theory of Zitterbewegung. *Phys. Rev. B* **81**, 121417(R) (2010).
44. L. Qifeng, Y. Yonghong, D. Jinming, Zitterbewegung in the honeycomb photonic lattice. *Opt. Lett.* **36**, 2513-2515 (2011).
45. J. Cserti, G. Dávid, Unified description ofZitterbewegungfor spintronic, graphene, and superconducting systems. *Phys. Rev. B* **74**, 172305 (2006).
46. J. Cserti, G. Dávid, Relation between Zitterbewegung and the charge conductivity, Berry curvature, and the Chern number of multiband systems. *Phys. Rev. B* **82**, 201405 (2010).
47. W. Ye, Y. Liu, J. Liu, S. A. R. Horsley, S. Wen, S. Zhang, Photonic Hall effect and helical Zitterbewegung in a synthetic Weyl system. *Light.: Sci. Appl.* **8**, 49 (2019).

# Valley-dependent wavepacket self-rotation and Zitterbewegung
# in symmetry-broken honeycomb lattices

Xiuying Liu[1+], Frane Lunić[2+], Daohong Song[1,3], Zhixuan Dai[1], Shiqi Xia[1], Liqin Tang[1],
Jingjun Xu[1,3], Zhigang Chen[1,3,4], and Hrvoje Buljan[1,2],

[1]The MOE Key Laboratory of Weak-Light Nonlinear Photonics, TEDA Applied Physics Institute and School of Physics, Nankai University, Tianjin 300457, China
[2]Department of Physics, Faculty of Science, University of Zagreb, Bijenička cesta 32, 10000 Zagreb, Croatia
[3]Collaborative Innovation Center of Extreme Optics, Shanxi University, Taiyuan, Shanxi 030006, People's Republic of China
[4]Department of Physics and Astronomy, San Francisco State University, San Francisco, California 94132, USA
songdaohong@nankai.edu.cn, hbuljan@phy.hr, zgchen@nankai.edu.cn
[+]These authors made equal contributions.

# Supplementary Material:

## Theory

**Dynamics via expansion in eigenmodes**

Here we theoretically calculate the dynamics of wavepackets in our linear photonic lattices by expanding the initial wavepacket (at $z=0$) into eigenmodes of the lattice. We present details of the calculation here, and only the key expressions are included in the main text. We consider graphene-like honeycomb lattices (HCLs) with a mass term, where the effective ("low-energy") Hamiltonian (in $k$-space) is (1)

$$H = \sigma_x k_x + \sigma_y k_y + \sigma_z m = \begin{pmatrix} m & k_x - ik_y \\ k_x + ik_y & -m \end{pmatrix} = \begin{pmatrix} m & ke^{-i\varphi_k} \\ ke^{i\varphi_k} & -m \end{pmatrix}, \quad (S1)$$

and the equation of motion is

$$i\frac{\partial \psi}{\partial z} = H\psi.$$

We use complex notation for the wavevector, $k_x + ik_y = ke^{i\varphi_k}$; the Pauli matrices $\boldsymbol{\sigma}$ are:

$$\sigma_x = \begin{pmatrix} 0 & 1 \\ 1 & 0 \end{pmatrix}, \sigma_y = \begin{pmatrix} 0 & -i \\ i & 0 \end{pmatrix}, \text{ and } \sigma_z = \begin{pmatrix} 1 & 0 \\ 0 & -1 \end{pmatrix}.$$

The eigenvalues (propagation constants) of the Hamiltonian (S1) are,

$$\beta_{nk} = n\sqrt{k^2 + m^2},$$

where $n = \pm 1$ is the band number, and the orthonormal eigenmodes are:

$$\psi_{n\mathbf{k}} = \frac{1}{\sqrt{N_{nk}}} \begin{pmatrix} \frac{m+\beta_{nk}}{ke^{i\varphi_k}} \\ 1 \end{pmatrix}, \quad N_{nk} = 2 + \frac{2m(m+\beta_{nk})}{k^2}.$$

The pseudospin up- and down- component corresponds to the excitation of **A** and **B** sublattice of the HCL, respectively.

We are interested in the propagation of the beam from an azimuthally symmetric initial state given by

$$\psi(r, \varphi_r, z = 0) = \psi_0 \sqrt{I_0(r)},$$

where

$$\psi_0 = \begin{pmatrix} \cos\theta \, e^{i\alpha} \\ \sin\theta \end{pmatrix}$$

is the most general initial spinor in Bloch sphere coordinates. We should note here that Eq. (S1) correctly describes dynamics for excitations in the vicinity of the $K$-point; therefore, the wavepacket in this "low-energy" theory is azimuthally symmetric. In experiments and numerical simulations, only the envelope of the initial excitation has azimuthal symmetry, but the fine structure of the beam's intensity conforms to the triangular lattice.

We can rewrite the initial complex amplitude in momentum space as follows:

$$\psi(r, \varphi_r, z = 0) = \int d^2k \, \psi_0 f(k) e^{i\mathbf{k}\cdot\mathbf{r}},$$

where the function $f(k)$ depends on the transverse profile of the initial excitation, and the integral is taken over the whole $k$-space. This initial condition can be written via superposition of the eigenmodes of the HCL by using

$$\psi_0 = \sum_n c_{n\mathbf{k}} \psi_{n\mathbf{k}},$$

where the coefficients $c_{n\mathbf{k}}$ are given by $c_{n\mathbf{k}} = \psi_{n\mathbf{k}}^+ \psi_0$. This yields

$$\psi(r, \varphi_r, z = 0) = \sum_n \int d^2k \, c_{n\mathbf{k}} \psi_{n\mathbf{k}} f(k) e^{i\mathbf{k}\cdot\mathbf{r}}.$$

The scalar products of the initial wavepacket with the eigenmodes are easily evaluated,

$$c_{n\mathbf{k}} = \psi_{n,\mathbf{k}}^+ \psi_0 = \frac{1}{\sqrt{N_{nk}}} \left( \frac{m+\beta_{nk}}{k} \cos\theta \, e^{i\alpha} e^{i\varphi_k} + \sin\theta \right),$$

from which we obtain

$$c_{n\mathbf{k}} \psi_{n\mathbf{k}} = \frac{1}{N_{nk}} \begin{pmatrix} \frac{m+\beta_{nk}}{k} \left( \frac{m+\beta_{nk}}{k} \cos\theta \, e^{i\alpha} + \sin\theta \, e^{-i\varphi_k} \right) \\ \frac{m+\beta_{nk}}{k} \cos\theta \, e^{i\alpha} e^{i\varphi_k} + \sin\theta \end{pmatrix}.$$

Finally, we obtain the propagation of the wave packet from a azimuthally symmetric initial condition via

$$\psi(r, \varphi_r, z) = \begin{pmatrix} \psi_{\frac{1}{2}}(r, \varphi_r, z) \\ \psi_{-\frac{1}{2}}(r, \varphi_r, z) \end{pmatrix} = \sum_n \int d^2k \, c_{n\mathbf{k}} \psi_{n\mathbf{k}} f(k) e^{i\mathbf{k}\cdot\mathbf{r} - i\beta_{nk}z} =$$

$$\sum_n \int_0^\infty k\,dk \int_0^{2\pi} d\varphi_k \frac{1}{N_{nk}} \begin{pmatrix} \frac{m+\beta_{nk}}{k} \left( \frac{m+\beta_{nk}}{k} \cos\theta \, e^{i\alpha} + \sin\theta \, e^{-i\varphi_k} \right) \\ \frac{m+\beta_{nk}}{k} \cos\theta \, e^{i\alpha} e^{i\varphi_k} + \sin\theta \end{pmatrix} e^{ikr\cos(\varphi_k - \varphi_r) - i\beta_{nk}z}. \quad (S2)$$

The integrals over the azimuth angle, $\int_0^{2\pi} d\varphi_k$, are analytically evaluated as

$$\int_0^{2\pi} d\varphi_k \, e^{ikr\cos(\varphi_k-\varphi_r)} = 2\pi J_0(kr), \text{ and}$$

$$\int_0^{2\pi} d\varphi_k \, e^{\pm i\varphi_k} e^{ikr\cos(\varphi_k-\varphi_r)} = 2\pi i e^{\pm i\varphi_r} J_1(kr),$$

where $J_0$ and $J_1$ are the Bessel functions; the remaining integral over $k$ needs to be evaluated numerically. From this we find the mathematical structure of the complex amplitude of the electric field:

$$\psi(r,\varphi_r,z) = \begin{pmatrix} \psi_{\frac{1}{2}}(r,\varphi_r,z) \\ \psi_{-\frac{1}{2}}(r,\varphi_r,z) \end{pmatrix} = \begin{pmatrix} g_{\frac{1}{2},0}(r,z) + g_{\frac{1}{2},-1}(r,z)e^{-i\varphi_r} \\ g_{-\frac{1}{2},+1}(r,z)e^{i\varphi_r} + g_{-\frac{1}{2},0}(r,z) \end{pmatrix},$$

where the $g$-functions are readily related to the remaining integrals over $k$ in Eq. (S2). Up to this point the calculation is generally valid for any azimuthally symmetric initial condition and for any pseudospin excitation. Intensity obtained via Eq. (S2) is illustrated in Figs. 3(a-c) in the main text, and self-rotating helical intensity pattern is clearly observed in our calculations.

**Self-rotating intensity structure**

The self-rotating helical intensity structure in each sublattice (i.e., for each pseudospin component) can be understood as the interference of the vortex beam component and the normal (non-vortex) beam component (see main text). To show this, first we write the intensity of the whole beam as

$$I(r,\varphi_r,z) = \psi^+\psi = \left|\psi_{\frac{1}{2}}(r,\varphi_r,z)\right|^2 + \left|\psi_{-\frac{1}{2}}(r,\varphi_r,z)\right|^2;$$

the intensity in the pseudospin-up component is

$$\left|\psi_{\frac{1}{2}}(r,\varphi_r,z)\right|^2 = \left|g_{\frac{1}{2},0}(r,z)\right|^2 + \left|g_{\frac{1}{2},-1}(r,z)\right|^2 + 2\mathrm{Re}\left(g^*_{\frac{1}{2},0}(r,z) g_{\frac{1}{2},-1}(r,z) e^{-i\varphi_r}\right).$$

The last term describes interference between the vortex and the non-vortex terms which can be written as

$$2\left|g_{\frac{1}{2},0}(r,z)\right|\left|g_{\frac{1}{2},-1}(r,z)\right|\cos\left(-\mathrm{Arg}\left(g_{\frac{1}{2},0}(r,z)\right) + \mathrm{Arg}\left(g_{\frac{1}{2},-1}(r,z)\right) - \varphi_r\right),$$

where $\mathrm{Arg}\left(g_{\frac{1}{2},0}(r,z)\right)$ denotes the phase of the normal component, and $\mathrm{Arg}\left(g_{\frac{1}{2},-1}(r,z)\right)$ denotes the phase of the vortex component. It is evident that $\left|g_{\frac{1}{2},0}(r,z)\right|^2 + \left|g_{\frac{1}{2},-1}(r,z)\right|^2$ is azimuthally symmetric, and therefore only the interference term $2Re\left(g^*_{\frac{1}{2},0}(r,z)g_{\frac{1}{2},-1}(r,z)e^{-i\varphi_r}\right)$ can produce the observed spiraling helical intensity pattern.

The maxima (minima) of the interference term occur when the cosine term is 1 (-1). Thus, the maxima are implicitly given by

$$\varphi_r = -\text{Arg}\left(g_{\frac{1}{2},0}(r,z)\right) + \text{Arg}\left(g_{\frac{1}{2},-1}(r,z)\right) \text{ (modulo } 2\pi\text{)},$$

which is Eq. (6) in the main text. At a given $z$-propagation distance, the behavior of the function $h(r) = -\text{Arg}\, g_{\frac{1}{2},0}(r,z) + \text{Arg}\, g_{\frac{1}{2},-1}(r,z)$ determines the location of the maxima of the interference term in the $(r, \varphi_r)$-plane. If $h(r)$ is monotonically increasing (or decreasing) with $r$, the function implicitly given by $\varphi_r = h(r)$ is a spiral in the $(r, \varphi_r)$-plane; the spiral helicity depends on whether $h(r)$ decreases or increases. Thus, the self-rotating spiraling intensity pattern is a very robust feature of the dynamics observed in our system.

## Zitterbewegung

The self-rotating intensity pattern is closely related to the Zitterbewegung phenomenon as clarified in the main text. In order to analyze the Zitterbewegung effect in this system, we study the dynamics of the center of mass (COM) of the wavepacket given by

$$x_C(z) = \langle x \rangle = \int \psi^+(r,\varphi_r,z) x \psi(r,\varphi_r,z) da, \text{ and}$$

$$y_C(z) = \langle y \rangle = \int \psi^+(r,\varphi_r,z) y \psi(r,\varphi_r,z) da,$$

where $da = r dr d\varphi_r$ is the infinitesimal element for the area integral. Dynamics of the COM can be understood by observing

$$v_x = \frac{\partial \langle x \rangle}{\partial z} = \int \psi^+(r,\varphi_r,z) i[H,x] \psi(r,\varphi_r,z) da = \langle \sigma_x \rangle,$$

and equivalently

$$v_y = \frac{\partial \langle y \rangle}{\partial z} = \langle \sigma_y \rangle.$$

These are the well-known results from the Zitterbewegung theory, adopted here for our system, and numerically verified in Fig. 3 of the main text. To reveal the underlying mechanism behind the Zitterbewegung, we further explore

$$\frac{\partial v_x}{\partial z} = \int \psi^+(r,\varphi_r,z) i[H,\sigma_x] \psi(r,\varphi_r,z) da = i\langle [H,\sigma_x] \rangle = -i2\langle k_y \sigma_z \rangle + i2m\langle \sigma_y \rangle, \quad (S3)$$

and

$$\frac{\partial v_y}{\partial z} = i\langle [H,\sigma_y] \rangle = i2\langle k_x \sigma_z \rangle - i2m\langle \sigma_x \rangle. \quad (S4)$$

It is convenient to define the vectors

$$\mathbf{P} = \langle k_x\sigma_z\rangle\hat{x} + \langle k_y\sigma_z\rangle\hat{y}, \text{ and}$$

$$\mathbf{v}_C = v_x\hat{x} + v_y\hat{y},$$

and rewrite Eqs. (S3) and (S4) as

$$\frac{\partial \mathbf{v}_C}{\partial z} = -2\,\hat{z}\times\mathbf{P} + 2m\,\hat{z}\times\mathbf{v}_C. \tag{S5}$$

The second term in Eq. (S5) corresponds to Zitterbewegung oscillations with frequency $2m$, i.e., with frequency corresponding to the gap size as expected. Moreover, it is clear that the sign of $m$ sets the helicity of the oscillations, which clarifies dependence of the helicity on the index offset between the sublattices and hence the valley-dependence.

Let us discuss the first term in Eq. (S5). First, we discuss components of the vector $\mathbf{P}$; the $x$-component is

$$\langle k_x\sigma_z\rangle = \int \psi^*_{\frac{1}{2}}(r,\varphi_r,z)k_x\psi_{\frac{1}{2}}(r,\varphi_r,z)da - \int \psi^*_{-\frac{1}{2}}(r,\varphi_r,z)k_x\psi_{-\frac{1}{2}}(r,\varphi_r,z)da,$$

and analogously the $y$-component is

$$\langle k_y\sigma_z\rangle = \int \psi^*_{\frac{1}{2}}(r,\varphi_r,z)k_y\psi_{\frac{1}{2}}(r,\varphi_r,z)da - \int \psi^*_{-\frac{1}{2}}(r,\varphi_r,z)k_y\psi_{-\frac{1}{2}}(r,\varphi_r,z)da.$$

Components of the vector $\mathbf{P}$ are interpreted as the difference of the expectation value of the momentum between the pseudospin-up and the pseudospin-down components, that is, the difference of the momentum between the two sublattices. The acceleration of the center of the beam in the $x$-direction is proportional to $P_y$, which can be therefore interpreted as a pseudo-force exerted in the center of the beam, which is of the curl-type.

**Zitterbewegung and the Berry connection**

The connection between the Zitterbewegung effect and the Berry phase has been addressed previously (2,3). Here we provide a thorough discussion on the relation between these phenomena in our system (the outline of the discussion can be applied in other systems as well). First, we rewrite the third term in Eq. (S2) in a slightly different form:

$$\psi(r,\varphi_r,z) = \int d^2k \left(\sum_n c_{n\mathbf{k}} f(k)\psi_{n\mathbf{k}} e^{-i\beta_{nk}z}\right) e^{i\mathbf{k}\cdot\mathbf{r}}.$$

Second, we point out that the dynamics of COM can be calculated in the momentum space as follows:

$$x_C(z) = \langle x\rangle = i\int \left(\sum_m c_{m\mathbf{k}} f(k)\psi_{m\mathbf{k}} e^{-i\beta_{mk}z}\right)^+ \frac{\partial}{\partial k_x}\left(\sum_n c_{n\mathbf{k}} f(k)\psi_{n\mathbf{k}} e^{-i\beta_{nk}z}\right) d^2k,$$

and

$$y_C(z) = \langle y\rangle = i\int \left(\sum_m c_{m\mathbf{k}} f(k)\psi_{m\mathbf{k}} e^{-i\beta_{mk}z}\right)^+ \frac{\partial}{\partial k_y}\left(\sum_n c_{n\mathbf{k}} f(k)\psi_{n\mathbf{k}} e^{-i\beta_{nk}z}\right) d^2k.$$

These equations allow us to explore relation between the Zitterbewegung effect and the Berry connection which is for the $n$th band ($n \in \{-1,1\}$) defined as $A_{n,x}(\mathbf{k}) = i\psi_{n\mathbf{k}}^+ \frac{\partial}{\partial k_x}\psi_{n\mathbf{k}}$, and $A_{n,y}(\mathbf{k}) = i\psi_{n\mathbf{k}}^+ \frac{\partial}{\partial k_y}\psi_{n\mathbf{k}}$. Next, we rewrite the expectation value $\langle x \rangle$ as follows:

$$\langle x \rangle = i \int \sum_{m,n} \left(c_{m\mathbf{k}} f(k) \psi_{m\mathbf{k}} e^{-i\beta_{mk}z}\right)^+ \frac{\partial}{\partial k_x}\left(c_{n\mathbf{k}} f(k) \psi_{n\mathbf{k}} e^{-i\beta_{nk}z}\right) d^2k =$$
$$i \int \sum_{m,n} (c_{m\mathbf{k}} f(k))^* \frac{\partial}{\partial k_x}(c_{n\mathbf{k}} f(k))(\psi_{m\mathbf{k}}^+ \psi_{n\mathbf{k}})(e^{i\beta_{mk}z} e^{-i\beta_{nk}z}) d^2k +$$
$$i \int \sum_{m,n} (c_{m\mathbf{k}} f(k))^* (c_{n\mathbf{k}} f(k))(\psi_{m\mathbf{k}}^+ \psi_{n\mathbf{k}})(e^{i\beta_{mk}z} \frac{\partial}{\partial k_x} e^{-i\beta_{nk}z}) d^2k +$$
$$\int \sum_{m,n} (c_{m\mathbf{k}} f(k))^* (c_{n\mathbf{k}} f(k))(i\psi_{m\mathbf{k}}^+ \frac{\partial}{\partial k_x} \psi_{n\mathbf{k}})(e^{i\beta_{mk}z} e^{-i\beta_{nk}z}) d^2k.$$

The first and the second terms are simplified because $\psi_{m\mathbf{k}}^+ \psi_{n\mathbf{k}} = \delta_{mn}$, whereas the third term obviously contains the Berry connections $A_{n,x}(\mathbf{k})$ and $A_{n,y}(\mathbf{k})$. Therefore, we can write:

$$\langle x \rangle = i \int \sum_n (c_{n\mathbf{k}} f(k))^* \frac{\partial}{\partial k_x}(c_{n\mathbf{k}} f(k)) d^2k + z \int \sum_n (c_{n\mathbf{k}} f(k))^* (c_{n\mathbf{k}} f(k)) \frac{\partial \beta_{nk}}{\partial k_x} d^2k +$$
$$\int \sum_n (c_{n\mathbf{k}} f(k))^* (c_{n\mathbf{k}} f(k)) A_{n,x}(\mathbf{k}) d^2k + \int (c_{1\mathbf{k}} f(k))^* (c_{-1\mathbf{k}} f(k))(i\psi_{1\mathbf{k}}^+ \frac{\partial}{\partial k_x} \psi_{-1\mathbf{k}}) e^{i2\beta_{1k}z} d^2k +$$
$$\int (c_{-1\mathbf{k}} f(k))^* (c_{1\mathbf{k}} f(k))(i\psi_{-1\mathbf{k}}^+ \frac{\partial}{\partial k_x} \psi_{1\mathbf{k}}) e^{-i2\beta_{1k}z} d^2k.$$

Let us examine this expression term by term. The first term is the center of the beam at $z = 0$, which follows from $\sum_n |c_{n\mathbf{k}}|^2 = 1$,

$$i \int \sum_n (c_{n\mathbf{k}} f(k))^* \frac{\partial}{\partial k_x}(c_{n\mathbf{k}} f(k)) d^2k = i \int f^*(k) \frac{\partial}{\partial k_x} f(k) d^2k = \langle x(0) \rangle.$$

The second term describes the drift of the center of the beam:

$$z \int \sum_n |c_{n\mathbf{k}} f(k)|^2 \frac{\partial \beta_{nk}}{\partial k_x} d^2k.$$

The third term is $z$-independent, and it contains an integral over weighted Berry connections in $k$-space; the weights depend on the initial conditions via $c_{n\mathbf{k}} f(k)$:

$$\int \sum_n |c_{n\mathbf{k}} f(k)|^2 A_{n,x}(\mathbf{k}) d^2k.$$

Finally, the fourth and the fifth terms correspond to the Zitterbewegung oscillations. Note that the Zitterbewegung oscillations are non-vanishing only if the interband Berry-type matrix elements $i\psi_{1\mathbf{k}}^+ \frac{\partial}{\partial k_x} \psi_{-1\mathbf{k}}$ and $i\psi_{-1\mathbf{k}}^+ \frac{\partial}{\partial k_x} \psi_{1\mathbf{k}}$ are nonvanishing. A fully equivalent analysis can be made for $\langle y \rangle$ but we omit that here.

Thus, we conclude that the nonvanishing interband Berry-type matrix elements are essential for the Zitterbewegung oscillations to occur, however, the Berry connection matrix elements are not present in the Zitterbewegung oscillation terms. In the aftermath, this is somewhat expected because Zitterbewegung oscillations were originally understood as oscillations between the positive and the negative energy states. Hence the Berry-type matrix elements between the positive and the negative energy states are essential.

It is important to note, however, that if the $k$-space eigenmodes $\psi_{n\mathbf{k}}$ were k-independent, then the Berry connection should be zero, and simultaneously the Zitterbewegung oscillations should be absent. An example of a system with 2-bands where both the Berry connection and Zitterbewegung are absent is associated with the following Hamiltonian:

$$H = \sigma_z \sqrt{k^2 + m^2},$$

where $\psi_{-1\mathbf{k}} = \begin{pmatrix} 0 \\ 1 \end{pmatrix}$, $\psi_{1\mathbf{k}} = \begin{pmatrix} 1 \\ 0 \end{pmatrix}$, and $\beta_{nk} = n\sqrt{k^2 + m^2}$. This example corroborates our finding that in systems with the non-vanishing Berry connection, the interband Berry-type matrix elements which are crucial for the Zitterbewegung are expected to be non-vanishing as well.

**References:**

(1) A. H. Castro Neto, F. Guinea, N. M. R. Peres, K. S. Novoselov, and A. K. Geim, *The electronic properties of graphene*, Rev. Mod. Phys. 81, 109 (2009).

(2) J. Y. Vaishnav and C. W. Clark, *Observing Zitterbewegung with Ultracold Atoms*, Phys. Rev. Lett. **100**, 153002 (2008).

(3) Cserti, J. and G. Dávid, *Relation between Zitterbewegung and the charge conductivity, Berry curvature, and the Chern number of multiband systems.* Physical Review B **82**, 201405 (2010).